\documentclass[10pt,conference,compsocconf]{IEEEtran}
\usepackage{amsmath,amsthm}
\usepackage{amssymb}
\usepackage{algorithmic,algorithm}
\usepackage{multirow}
\usepackage{threeparttable}
\usepackage{array}
\usepackage{graphicx,color}
\usepackage{subfigure}
\usepackage{listings}
\usepackage{caption}
\usepackage{pifont}
\usepackage{url}
\usepackage{array}
\usepackage[bookmarks=false]{hyperref}
\usepackage[skins,listings]{tcolorbox}
\usepackage{amsfonts}
\usepackage{float}
\usepackage{framed}
\usepackage{soul}
\usepackage{xcolor}

\newtcolorbox{codebox}[1]{
colback=white,
colframe=black,
fonttitle=\sffamily\bfseries\normalsize,
enhanced,
boxsep=0pt,
left=0pt,
right=0pt,
top=2pt,
title=#1}

\definecolor{myred}{rgb}{1, 0.0, 0.22}
\makeatletter
\renewcommand{\ALG@name}{Conjecture}

\newtheorem{P1:fine}{Definition}[]

\newcounter{prove}[section]
\newenvironment{prove}[2][]
{\refstepcounter{prove}
   \noindent\textbf{\texttt{Property~\theprove .(#2) #1}} \rmfamily}

\newcounter{evidence}[section]
\newenvironment{evidence}[1][]
{\refstepcounter{evidence}\par\medskip
   $\textit{Evidence~\theevidence. #1}$ \rmfamily}{\medskip}

\newcounter{definition}[section]
\newenvironment{definition}[1][]
{\refstepcounter{definition}\par\medskip
   \textit{Definition~\thedefinition. #1} \rmfamily}{\medskip}

\newcounter{observation}[section]
\newenvironment{observation}[1][]
{\refstepcounter{observation}\par\medskip
   \textit{Observation~\theobservation. #1} \rmfamily}{\medskip}

\newcounter{evidenceiii}[section]
\renewcommand{\theevidenceiii}{\roman{evidenceiii}}
\newenvironment{evidenceiii}[1][]
{\refstepcounter{evidenceiii}\par\medskip
   $\textit{Evidence~\theevidenceiii. #1}$ \rmfamily}{\medskip}

\makeatother
\newcolumntype{x}[1]{>{\centering\let\newline\\\arraybackslash\hspace{0pt}}p{#1}}
\definecolor{mygray}{gray}{0.2}

\begin{document}
\title{\textsc{Gpt} Conjecture: Understanding the Trade-offs between Granularity, Performance and Timeliness in Control-Flow Integrity}  
\author{\IEEEauthorblockN{Zhilong Wang,
Peng Liu}
\IEEEauthorblockA{College of Information Sciences and Technology\\
The Pennsylvania State University, USA\\
zzw169@psu.edu, pliu@ist.psu.edu}
}

\maketitle

\begin{abstract}
Performance/security trade-off is widely noticed in CFI research, however, we observe that not every CFI 
scheme is subject to the trade-off. 
Motivated by the key observation, we ask three questions. 
Although the three questions probably cannot be directly answered, they are inspiring. 
We find that a deeper understanding of the nature of the trade-off will help answer the three questions. 
Accordingly, we proposed the \textsc{Gpt} conjecture to pinpoint the trade-off in designing CFI schemes, 
which says that at most two out of three properties (fine granularity, acceptable performance, and preventive protection) could be achieved.
\end{abstract}

\section{Introduction}



Along with the increased complexity of software, 
it becomes harder for the developers to ensure execution correctness in their software products, 
especially in those developed by the low-level programming languages, such as C/C++.
A substantial amount of execution in-correctness is caused by the exploitation of software vulnerabilities in the real world.
Softwares inevitably contain a wide variety of vulnerabilities, 
opening a window for attacks to compromise the system.
Attackers have developed a series of attack methods, such as shellcode injection\cite{Erickson2008Hacking},
return-to-libc\cite{wojtczuk2001advanced}, ROP\cite{shacham2007geometry} and so on, 
to exploit all kinds of vulnerabilities, e.g., buffer overflow, format string, use-after-free, and so on~\cite{szekeres2013sok}.
Among all kinds of attacks, the control-flow hijacking attack is the most dangerous one, 
because it allows the attacker to control the program's execution, execute arbitrary malicious code
and attain Turing-complete operation\cite{shacham2007geometry}.
To mitigate the threats, many defense mechanisms, 
such as stack smashing protector (SSP)\cite{cowan1998stackguard}, 
address space layout randomization (ASLR)\cite{aslr}, data execution prevention (DEP)\cite{dep} and so on,
have been put forward by researchers and applied in the real world software products.

Among all the defense techniques, security schemes based on the concept of control-flow integrity (CFI) 
have attracted many researchers' attention because of its simplicity to implement,
effectiveness to cope with the full spectrum of control-flow hijacking attacks, 
and flexibility to trade between security and efficiency. 
CFI schemes guarantee the correctness of the program by dynamically checking the control-flow transfer
and confining the target address to a legal set. 

Since CFI was introduced by Abadi et al. in 2005 \cite{Abadi}, 
many researchers afterward were dedicated to enhance its runtime performance, security, scalability, compatibility and so on.
According to mainstream taxonomy, most CFI schemes can be clarified into two categories: 
fine-grained CFI schemes that provide more security guarantee, and coarse-grained CFI schemes that attain higher runtime performance.
However, both fine-grained and coarse-grained CFI schemes have noticeable limitations that have not been addressed yet. 
As shown in previous survey papers~\cite{burow2017control}, 
lightweight CFI schemes can not fully prevent sophisticated code reuse attack.
The adversary’s attacking strategy is to search large gadgets chain whose starting addresses are allowed in a rough control-flow graph that
coarse-grained CFI schemes adopted~\cite{goktas2014out,passcfi}. 
Precise CFI schemes usually suffer from unacceptable runtime overhead.
Hence, it is widely believed ``performance/security trade-off'' 
exists between runtime overhead and security in different CFI schemes~\cite{burow2017control,236352}.

However, we observe that not every CFI scheme is subject to the trade-off between performance and security. 
In fact, several CFI schemes are ``immunized'' from doing such a trade-off.
For instance, $\pi$CFI designed by Niu et al. achieves fine-grained security with a runtime overhead of 3.2\% on average,
which is fairly low and acceptable~\cite{niu2015per}.
Victor et al. proposed a context-sensitive CFI scheme that achieves 
stronger security than conventional fine-grained ones with an overhead of less than some of the coarse-grained ones~\cite{van2015practical}.

{\bf Key Observation.} The trade-off between performance and security does \emph{not} universally exist in meaningful CFI schemes. 
This intriguing observation motivates us to ask three questions: \ding{202} does trade-off really exist in different CFI schemes? 
\ding{203} if trade-off do exist, How do previous works comply with it? 
\ding{204} how can it inspire future research?

Although the questions probably cannot be directly answered, they are inspiring. 
On the other hand, we find that a deeper understanding of the nature of the trade-off will help answer these questions.
Accordingly, we propose the \textsc{Gpt} conjecture to pinpoint general trade-offs in CFI schemes:
the impossibility of guaranteeing both fine granularity and acceptable performance in a Just-In-Time CFI scheme.
We analyze its rationality through empirical study\textemdash surveying a series of representative CFI schemes and
showing how existing CFI schemes comply with our conjecture.
Finally, we give some recommendations for future researchers.
We believe that our conjecture will help researchers have a more clear understanding of internal relations among properties of CFI schemes, 
thereby, motivating future research in this area.

\section{Background}
When compiling source code written by low-level language (such as C or C++) into machine code,
the compiler emits control data~\cite{chen2005non} (data that are loaded to processor program counter at some point in program execution, 
e.g., return addresses and function pointers)
into the binary file without any protection. 
The security of control data depends on checks inserted by the programmer to enforce memory safety~\cite{nagarakatte2012practical}. 
Along with program execution, 
attacker's malicious tampering with control data through software vulnerabilities, 
such as buffer overflow, can transfer the program's control-flow to any executable address in process space.

Based on this observation, researchers invented CFI to protect programs against control-flow hijacking attacks
by checking programs' control data before loading them into the program counter (EIP/RIP register in x86/x64 architecture). 
CFI's strategy is to restrict the control-flow of a program 
to a pre-calculated CFG by checking indirect control-flow transfers at runtime~\cite{burow2017control}.
Generally, most of CFI schemes follow a mainstream that consists of two phases.

In phase one, an analyzer statically computes the program's control-flow graph (CFG).
CFG is a representation in graph form of all legitimate control-flow transfers (also being called branch) in program space. 
It consists of sets of nodes and directed edges.
Each node and edge denotes a basic block and a valid branch in the program respectively. 
For a comprehensive understanding, we refer the reader to the formal definition of CFG in work by Allen, et al.~\cite{Allen:1970}.

In phase two, a runtime control-flow checking (validation) component validates just fetched control data before each indirect-branching
according to the legitimate CFG generated in phase one~\footnote{Direct control-flow transfers do not load any control data, their target addresses/offsets are hard-coded in their instructions.}.
An indirect-branch can pass checking only if it can be matched to a corresponding edge in the CFG.
A failed validation will result in the process to terminate its execution and report an error.
In such a fashion, control-flow attacks which usually introduce out-of-range branch are extremely prohibited.
Researchers need to design efficient data structures to represent the CFG and enable runtime checking.

Despite its straightforward main idea, 
it is pretty challenging to design a CFI scheme with strong security, acceptable performance, high compatibility and so on~\cite{236352,burow2017control}.
Researchers have designed hundreds of CFI schemes to explore its potential in different perspectives.
The dominant difference of these various CFI schemes can be summarized into three aspects:
1) the precision of a CFG they employed. 2) the algorithm they designed to check indirect-branches. 
3) the time point checking algorithm was activated. 

\subsubsection{Precision of CFG Analyzer}
CFG can be obtained by analyzing the program's source code or binary code.
Like pointer analysis~\cite{pointeranalysis}, perfect CFG generation is can not be fully achieved yet in many situations~\cite{goktas2014out}.
By now researchers have adopted several types of methods (insensitive analysis, context-sensitive analysis, and path-sensitive analysis) in their CFG analyzer 
and achieve different precisions.
It is widely agreed that path-sensitive analysis is more precise than context-sensitive analysis, and
context-sensitive analysis is more precise than insensitive analysis~\cite{khedker2017data}.

\subsubsection{Algorithm to Enforce Checking}
\label{checkingalgorithm}
The efficiency of different CFI schemes is largely dependent on their algorithms to enforce validation, 
which is tightly combined with their data structure that represents the CFG and enables runtime checking. 
Researchers have designed different types of algorithms and data structures in different CFI schemes.
For example, the original CFI scheme proposed by the Abadi, et al. groups branch targets into different sets, assigns each set with a label,
and inlines labels into each jump targets, i.e., the basic block's in code. 
Based on this data structure, ``guard instructions'' are emitted before each indirect-branch instruction to compare its label with the one in target basic block~\cite{Abadi}.  
A mismatch indicates that the control data is corrupted, 
then the program's execution will be redirect to the error handling code accordingly.

$\pi$CFI~\cite{niu2015per} and MCFI~\cite{mcfi} by Niu, et al. adopts two ID tables, namely Bary and Tary, to store target program's CFG.
In essence, Bary table and Tary table are hashmaps which can efficiently map indirect-branch points and target basic blocks to their corresponding IDs.
Specifically, the Tary table is an array of IDs indexed by code addresses, mapping target basic block to their corresponding IDs.
The Bary table uses a similar design, mapping indirect-branch points to their corresponding IDs.
Two tables enable efficient ID look-ups and 
a indirect-branch is checked by comparing the IDs of branch point and target. 

\subsubsection{Just-In-Time Checking vs. Lazy Checking}
\label{sec:offline}
Another difference among CFI schemes is how they schedule their checking operations.
Most CFI schemes check the target address before indirect-branch occurs (we define it as a \texttt{Just-In-Time checking}).
While, to achieve better performance, some works log each indirect-branches at runtime and  
check them by employing another accompanying thread~\cite{huenforcing,203656,griffin,van2015practical,203656} (we define it as \texttt{Lazy checking}).
For example, PITTYPAT~\cite{203656} enforces path-sensitive CFI by maintaining a ``shadow'' execution/analyzer, 
running concurrently with the protected process and checks its finished indirect-branches. 
Such a non-intrusive checking does not disturb the normal execution of the monitored process, hence achieves path-sensitive CFI with practical runtime overhead.
\section{Conjecture}

%
This section aims to answer Question\ding{202} and Question\ding{203}.
We observe that some terms, such as coase-grained/fine-grained, have not been clearly defined. 
Before introducing the \textsc{Gpt} conjecture, 
let us give a more precise definition of the terms and concepts that will be used throughout 
the paper.
Then we propose the \textsc{Gpt} conjecture which helps to answer Question\ding{202}. 
At last, some evidence is collected from an empirical study to answer the Question\ding{203}.
\subsection{Terminology}
\label{sec:terms}

\begin{prove}{Granularity}

  Suppose a program has $n$ indirect branch instructions.
  Let $\mathbb{Z}_i$~\footnote{It is computed through mainstream insensitive control flow analysis. We admit the inaccuracy due to the difficulty of the pointer analysis.}
   denote the set of valid successors (basic blocks) of the $i$-th indirect branch instruction, 
  and $\mathcal{S}$ denote the set of all successor sets, namely,
  \begin{equation}
    \label{theo:eq1}
    \mathcal{S}=\left\{\mathbb{Z}_i: 1 \leq i \leq n\right\}
  \end{equation}

  For a CFI scheme,
  let $\mathbb{C}_i$ denote the checking set which is defined by the scheme 
  and assigned to the $i$-th indirect branch instruction, then used to check the branch's target at runtime.
  Only the elements in $\mathbb{C}_i$ are valid successors authorized by the CFI schemes
  that the $i$-th branch instruction could jump to.

  \begin{definition} \label{def:fine}For arbitrary two sets $\mathbb{Z}_i$, $\mathbb{Z}_j$ from $\mathcal{S}$, 
  satisfying $\mathbb{Z}_i \cap \mathbb{Z}_j \neq \varnothing \lor \mathbb{Z}_i \neq \mathbb{Z}_j$,
  as long as the CFI scheme merges $\mathbb{Z}_i$, $\mathbb{Z}_j$ when define its $\mathbb{C}_i$ or $\mathbb{C}_j$, namely,
  \begin{equation}
    \label{theo:eq2}
    \mathbb{Z}_i \cup \mathbb{Z}_j  \subset \mathbb{C}_i \lor
    \mathbb{Z}_i \cup \mathbb{Z}_j  \subset \mathbb{C}_j
  \end{equation}
  we define this scheme as a \texttt{coase-grained CFI scheme}.
  Otherwise, we define it as a \texttt{fine-grained CFI scheme}.
  This definition enables us to determinate the \texttt{granularity} property of CFI schemes.
  \end{definition}

  \textsc{Remark 1.} According to Definition~\ref{def:fine}, both
  context-sensitive and path-sensitive CFI schemes belong to \texttt{fine-grained CFI scheme}.
  In essence, they reduce the size of their checking set $\mathbb{C}_i$ for $i \in[1, n]$ based on context-sensitive or path-sensitive pointer analysis.
  Their protection is generally considered to be more powerful than that of insensitive \texttt{fine-grained CFI scheme}.

  \textsc{Remark 2.} Note that CFI schemes~\cite{mashtizadeh2015ccfi,8440029} which adopt pointer encryption approach should be 
  classified as \texttt{coase-grained CFI scheme}.
  They cannot fully prevent code reuse attack because of two noticeable drawbacks. 
  As discussed in Cryptographically Enforced Control Flow Integrity (CCFI)~\cite{mashtizadeh2015ccfi},
  it is still possible to replace 
  the current encrypted pointer with another one from the program space 
  and potentially disrupt control flow.
  The other drawback is that these schemes suffer from key leakage issues:
  the key can be infered by brute-force attack or known-plaintext attack~\cite{peng2006known},
  especially for schemes which adopt a linear encryption/decryption method (XOR)~\cite{8440029}. 

  \textsc{Remark 3.} We remark that schemes that only provide partial protection\textemdash 
  protecting subset of indirect branches in program space\textemdash 
  belong to \texttt{coase-grained CFI scheme}. 
  For instance, \texttt{vfGuard}~\cite{prakash2015vfguard}, VTV~\cite{tice2014enforcing}, and \textsc{SafeDispatch}~\cite{jang2014safedispatch} 
  only achieve strict protection for virtual function calls in COTS binaries;
  
\end{prove}

\begin{prove}{Performance}

  \begin{evidence}
    As discussed in many papers~\cite{szekeres2013sok,burow2017control,microsoft},
    runtime performance is one of the most important determinants of whether a defense technique will be adopted by industry.
    Generally, to get adopted by industry, a defense technique should introduce less than 5\% average overhead, such as StackGurad, ASLR, and DEP.
    Techniques incuring an overhead larger than 10\% do not tend to gain wide adoption in production environments. 
    Accordingly, the threshold should lie between 5\%-10\%.
  \end{evidence}
  \begin{evidence}
    Other than runtime performance, space performance is another important index to measure a scheme.
    Program's runtime memory consumption consists of four aspects, i.e., code, global data, heap, and stack. 
    Different programs have different ratios in four aspects, 
    and a defense technique commonly increases memory consumption in one or more aspects. 
    We observe that shadow based protections like shadow stack~\cite{shadowstack}, shadow memory~\cite{newsome2005dynamic} and shadow processing~\cite{patil1995efficient},
    that double memory consumption in one or more aspects are unlikely to be deployed in practice.
  \end{evidence}

\begin{definition}
  Conservatively, we define a runtime overhead of less than 10\% and a space overhead of less than 100\% (in any of aforementioned four aspects) as an \texttt{acceptable performance}.
  Otherwise, it is an \texttt{unacceptable performance}. 
  This definition enables us to determinate the \texttt{performance} property of CFI schemes.
\end{definition}
 
\end{prove}

\begin{prove}{Timeliness}

  \begin{observation}
    Whereas the term ``integrity'' in the context of CFI implies that it can prevent the attacks~\cite{Abadi},
    some of the CFI schemes do not hit the mark.
    To achieve higher efficiency, 
    some CFI schemes as mentioned in Section~\ref{sec:offline} adopted a \texttt{lazy checking} mechanism, which 
    checks programs' control-flow following the program‘s execution rather than before each indirect branching.
    Generally, they log the program's runtime control-flow transfer along with its execution,
    then check the control-flow offline or through an accompanying thread.
    In these designs, a sliding window exists between the program's control-flow transfer and checking.
    The attacker can compromise the system without being perceived in the sliding window, which means this kind of CFI cannot protect software against such attacks.
    \end{observation}
    \begin{definition}
      \label{def:timeliness}
      We regard that the aforementioned design of CFI schemes provides less protection than CFI schemes that perform \texttt{Just-In-Time checking}.
      We define protection capability powered by \texttt{lazy checking} schemes 
      as \texttt{detective protection},
      the others that powered by \texttt{Just-In-Time checking} as \texttt{preventive protection}.
      This definition enables us to determinate the property \texttt{timeliness} of CFI schemes.
    \end{definition}
    \end{prove}

\subsection{The Proposed Conjecture}
\label{def:conj}
\definecolor{shadecolor}{gray}{0.8}
\begin{shaded*}
  \noindent\textbf{\texttt{\textsc{Gpt} Conjecture:}}
  A control-flow integrity scheme can have at most two out of three properties: \vspace{0.6mm}\\
    \hspace*{3mm}\textbf{\texttt{P1.}} \texttt{Fine granularity}\\
    \hspace*{3mm}\textbf{\texttt{P2.}} \texttt{Acceptable performance}\\
    \hspace*{3mm}\textbf{\texttt{P3.}} \texttt{Preventive protection}
\end{shaded*}
\begin{table}[h]
  \centering  
  \captionsetup{justification=centering}
  \caption{Reflection of \textsc{Gpt} conjecture in 32 control-flow integrity schemes.}
  \begin{threeparttable}
\makeatletter
\newcommand{\widenhline}{%
    \noalign {\ifnum 0=`}\fi \hrule height 0.5pt
    \futurelet \reserved@a \@xhline
}
\newcolumntype{I}{!{\vrule width 1pt}}
\makeatother
\centering
\small
\begin{tabular}{cIc|x{1cm}|c|x{1cm}}  
\hline  
\hline  
Schemes                 &  & \textbf{\texttt{P1}}\tnote{1}  & \textbf{\texttt{P2}} & \textbf{\texttt{P3}}   \\ 
\hline  
\hline  
CFIXX~\cite{burow2018cfixx} &           & {\color{myred}\ding{55}}        & 4.98\%         &  \ding{51}\\ 
\hline  
\textsc{Reins}~\cite{wartell2012securing} &           & {\color{myred}\ding{55}}        & 2.40\%         &  \ding{51}\\ 
\hline  
\texttt{vfGuard}~\cite{prakash2015vfguard} &           & {\color{myred}\ding{55}}        & 18.30\%         &  \ding{51}\\ 
\hline  
VTV~\cite{tice2014enforcing} &           & {\color{myred}\ding{55}}        & 9.60\%         &  \ding{51}\\ 
\hline  
LLVM-CFI~\cite{llvmcfi}&             & {\color{myred}\ding{55}}       & 1.10\%          &  \ding{51}\\  
\hline  
VTI~\cite{bounov2016protecting}&           & {\color{myred}\ding{55}}       & 0.50\%         &  \ding{51}\\ 
\hline  
CFGuard~\cite{cfguard}&           & {\color{myred}\ding{55}}       & 2.30\%         &  \ding{51}\\  
\hline  
IFCC~\cite{tice2014enforcing} &          & {\color{myred}\ding{55}}      & -0.30\%         &  \ding{51} \\ 
\hline   
ROPecker~\cite{ropecker} &          & {\color{myred}\ding{55}}       & 2.60\%         &  \ding{51} \\ 
\hline  
bin-CFI~\cite{zhang2013control} &         & {\color{myred}\ding{55}}       & 8.50\%         &  \ding{51} \\  
\hline  
ROPGuard~\cite{fratric2012ropguard} &         & {\color{myred}\ding{55}}      & 0.48\%        &  \ding{51} \\
\hline   
SafeDispatch~\cite{jang2014safedispatch} &           & {\color{myred}\ding{55}}       & 2.00\%         &  \ding{51} \\ 
\hline  
CCFIR~\cite{zhang2013practical}&        & {\color{myred}\ding{55}}       & 2.08\%         &  \ding{51}\\  
\hline 
kBouncer~\cite{pappas2013transparent}&           & {\color{myred}\ding{55}}      & 4.00\%         &  \ding{51} \\ 
\hline  
OCFI~\cite{mohan2015opaque} &           & {\color{myred}\ding{55}}       & 4.70\%         &  \ding{51} \\ 
\hline 
CFIMon~\cite{xia2012cfimon} &           & {\color{myred}\ding{55}}       & 6.10\%         &  \ding{51} \\ 
\hline  
$\tau$CFI~\cite{muntean2018tau} &           & {\color{myred}\ding{55}}       & 2.89\%         &  \ding{51} \\ 
\hline 
HCIC~\cite{8440029}&           & {\color{myred}\ding{55}}        & 0.95\%            &  \ding{51} \\  
\hline 
RAGuard~\cite{Zhang:2017}&           & {\color{myred}\ding{55}}        & 1.86\%            &  \ding{51} \\  
\hline 
HyperSafe~\cite{5504800}&           & {\color{myred}\ding{55}}        & 5.00\%            &  \ding{51} \\  
\hline 
\texttt{BinCC}~\cite{5504800}&           & {\color{myred}\ding{55}}        & 4.00\%            &  \ding{51} \\  
\hline  
CCFI~\cite{mashtizadeh2015ccfi}&           & {\color{myred}\ding{55}}        & {\color{myred}52.00\%}         &  \ding{51} \\  
\hline   
\textsc{KCoFI}~\cite{6956571} &           & {\color{myred}\ding{55}}        & {\color{myred}13.00\%}        &  \ding{51}\\ 
\hline  
Original CFI~\cite{abadi2009control}&            & \ding{51}      & {\color{myred}16.00\%}         &  \ding{51}\\ 
\hline 
Lockdown~\cite{payer2015fine}&           & \ding{51}       & {\color{myred}20.00\%}         &  \ding{51} \\ 
\hline  
MCFI~\cite{mcfi}        &           & \ding{51}             & 5.00\% \& {\color{myred}4GB}         &  \ding{51} \\
\hline 
$\pi$CFI~\cite{niu2015per}&           & \ding{51}            & 3.20\% \& {\color{myred}4GB}        &  \ding{51} \\   
\hline 
\textsc{Griffin}~\cite{griffin} &  {\scriptsize H\tnote{2}}        & \ding{51}          & {\color{myred}11.90\%}        &  {\color{myred}\ding{55}} \\ 
\hline  
\textsc{PittyPat}~\cite{203656} & {\scriptsize P}, {\scriptsize H}  & \ding{51}       & {\color{myred}12.73\%}         &  {\color{myred}\ding{55}} \\ 
\hline 
\textsc{ECFI}~\cite{Abbasi:2017} &         & \ding{51}          &    1.50\%        &  {\color{myred}\ding{55}} \\ 
\hline 
$\mu$CFI~\cite{huenforcing}& {\scriptsize C}, {\scriptsize H}        & \ding{51}       & 10.00\%             &  {\color{myred}\ding{55}} \\ 
\hline 
PathArmor~\cite{van2015practical}& {\scriptsize C} & \ding{51}         & 3.00\%         &  {\color{myred}\ding{55}} \\ 
\hline 
\hline   
\end{tabular} 
    \begin{tablenotes}
        \scriptsize
        \item[1] If a CFI scheme supports different security levels, e.g. having both \texttt{coase-grained} and \texttt{fine-grained} versions, we focus on its most secure version.
        \item[2] `H', `P' and `C' denote hardware-assisted CFI scheme, path sensitive CFI scheme, and context sensitive CFI scheme, respectively.
    \end{tablenotes}
\end{threeparttable}
\label{tab:comp}
\end{table}

\subsection{Some Evidence of the \textsc{Gpt} Conjecture}

In this section, we will reflect on our conjecture through several pieces of evidence. 
To verify the rationality of our conjecture, 
  we conduct an empirical study on 32 representative works, 
  and show the results in Table~\ref{tab:comp}.
  Three columns (\textbf{\texttt{P1}}, \textbf{\texttt{P2}} and \textbf{\texttt{P3}}) 
  in the table display three properties respectively as we define in Section~\ref{sec:terms}.  
  \textbf{\texttt{P1}} column denotes the granularity\textemdash
  check-mark indicates a \texttt{fine-grained scheme} whileas cross-mark represents a \texttt{coase-grained scheme}.
  \textbf{\texttt{P2}} column shows the performance overheads which are reported in corresponding papers. 
  Note that we prefer evaluation results which are based on 
  $\text{SPEC CPU}^ {\scriptsize{\textcircled{\tiny{R}}}}$2006 benchmarks~\cite{spec}.
  \textbf{\texttt{P3}} column labels whether a CFI scheme provides \texttt{preventive protection}.
  We label the data in each column with \emph{red} color when it fails to meet the requirement 
  defined in the conjecture. 
\begin{evidenceiii}
  It can be clearly seen in Table~\ref{tab:comp} that all CFI schemes we surveyed comply with our 
  conjecture\textemdash\textbf{no CFI schemes can achieve all three properties}.
  Also, some of unsophisticated schemes, such as \textsc{PittyPat}~\cite{203656} and \textsc{Griffin}~\cite{griffin}, 
  only achieve one properity, i.e., \texttt{fine granularity}. 
\end{evidenceiii}

\begin{evidenceiii}
\label{evi:space}
MCFI~\cite{mcfi} and $\pi$CFI developed by Niu, et al. achieve \texttt{fine granularity}
with acceptable runtime overheads, i.e., 3.2\% and 5.0\%, respectively. 
However, researchers did not realize that their better runtime overhead is achieved through sacrificing their space performance.
Even though they did not report their space overhead in their paper explicitly,
we can infer it in a reasonable manner.

As discussed in Section~\ref{checkingalgorithm}, both of two schemes adopt two tables, namely Bary and Tary, to support their runtime checking.
Accordingly, 1GB/4GB memory space on x86-32 and x86-64 operating system, respectively, need to be reserved in each process for the tables.
As stated by the author, ``On x86-32, memory segmentation is used, as in NaCl~\cite{5207638}. 
A 1GB segment is reserved for running the application code and another 1GB segment is reserved for the table region.
x86-64, however, does not support memory segmentation. 
Instead, memory writes are instrumented so that they are restricted to the [0, 4GB) memory region. 
Another 4GB memory region is reserved for tables.''
In view of the size of memory consumption of typical programs (mostly less than 1GB~\cite{spec}), 
their space overhead has already reached 100\% except for code bloat caused by extra no-op instructions inserted to enforce four-byte alignment on indirect-branch targets.
\end{evidenceiii}

\begin{evidenceiii}
  \label{hardwarecfi}
  \textsc{Griffin}~\cite{griffin} is a hardware-assisted CFI, 
   which leverages Intel PT to record control-flow of a monitored program. 
  It supports multiple types of CFI policies to enable flexible trade-offs between security and performance. 
  The \texttt{fine-grained} scheme incures an average of 11.9\% overhead.
  It leverages idle cores on a multi-core system for security checking 
  by having multiple worker threads to check runtime control-flow simultaneously. 
  In most of the time, it performs non-blocking checking which analyzes trace buffer of Intel PT whenever it becomes full;
  In a few cases when security-sensitive system calls are invoked, it performs blocking checking which stops the target thread until all the control transfers in the buffer have been checked.
  It can only provide the \texttt{detective protection} for software according to Definition~\ref{def:timeliness}.
  This case indicates that \textsc{Gpt} conjecture is applicable to hardware-assisted CFI schemes. 
\end{evidenceiii}

\begin{evidenceiii}
\textsc{PittyPat}~\cite{203656}, $\mu$CFI~\cite{huenforcing} and PathArmor~\cite{van2015practical}
are path/context sensitive CFI schemes which adopt path-sensitive or context-sensitive analysis to generate their CFG. 
However, path-sensitive and context-sensitive analysis is generally considered to be more time-consuming and space-consuming than insensitive analysis~\cite{khedker2017data}. 
We find that all three CFI schemes adopt two common features: 
\texttt{hard-assisted branch recording} and \texttt{lazy checking}.
Specifically, \textsc{PittyPat} and $\mu$CFI employ Intel PT\textemdash a brand new hardware feature in Intel CPUs\textemdash to efficiently record conditional and indirect branches taken by a program at runtime 
while PathArmor adopts Last Branch Record (LBR) registers available in Intel processors to monitor recently exercised control-flow transfers in an efficient way.
Their control-flow checking is achieved through accompanying threads. 
This case indicates that both path-sensitive and context-sensitive CFI schemes conform to the claim of \textsc{Gpt} conjecture.
\end{evidenceiii}

\textsc{Remark 4.}
Our observations indicate that the \textsc{Gpt} conjecture is universally applicable 
in all kinds of scenarios. 
Further, four pieces of evidence are not meant to be exhaustive and more evidence are easy to find.

\section{Implications of the \textsc{Gpt} Conjecture}
In this section, we will focus on answering Question \ding{204}: how can \textsc{Gpt} conjecture inspire future research?

First of all, \textsc{Gpt} conjecture illustrates the inherent trade-offs of three important properties (\texttt{fine granularity}, \texttt{acceptable performance}, and \texttt{preventive protection}) in CFI schemes.
It helps researchers to have a deeper understanding of the nature of CFI based protection. 
Accordingly, future researchers should make a necessary sacrifice before designing new CFI schemes.
In the broader context, \textsc{Gpt} conjecture provides insights into the feasible design space for CFI schemes, 
shedding some light on the manner in which algorithm designers and software engineers have circumvented the conjecture.

Second, for decades, security researchers have been focused on CFI scheme's runtime performance and made their best effort to improve it.
Evidence~\ref{evi:space} shows that in some cases, better runtime performance is achieved by sacrificing its space performance.
Just as Gerhard states, ``For some problems, we can reach an improved time complexity, 
but it seems that we have to pay for this with an exponential space complexity''~\cite{timeandspace}.
Therefore, performance evaluation in future research should not merely be limited to runtime performance and researchers should have a more comprehensive evaluation of their schemes. 

Third, Evidence~\ref{hardwarecfi} that even powerful hardware support cannot
eliminate the runtime overhead of \texttt{Just-In-Time} CFI schemes to an acceptable level, 
which implies that the challenge in the implementation of CFI 
cannot be solved only through engineering efforts,
instead, it may relate to computational complexity theory~\cite{comcomplexity}.
In a broader sense, we observe that indirect branching poses not only challenge in the security field,  
but also challenges to many others:
precise pointer analysis is NP-hard~\cite{pointernp};
indirect branch prediction is a performance-limiting factor for current computer systems~\cite{branchpredict}. 
Hence, \textsc{Gpt} conjecture implies the complexity of the CFI problem,
which deserves to be investigated through theoretical methods. 

At last, despite the inspiring implications that Gpt conjecture gives to us,
we admit that we still cannot prove the conjecture at this time. 

\section{Conclusion}
Control-flow integrity is a popular defence technique for detecting and defeating control-flow hijacking attacks.
Since its inception in the decade, researchers have put great efforts to explore its
potential regarding security, performance, compatibility and so on.
Even though performance/security trade-off is widely noticed in CFI research, 
we observe that not every CFI scheme is subject to it.
In this paper, we propose the \textsc{Gpt} conjecture to illustrate the general trade-offs in CFI schemes.
The conjecture points out the impossibility of guaranteeing both \texttt{fine granularity} and \texttt{acceptable performance} in a \texttt{Just-In-Time} CFI schemes. 
We have verified the rationality of our conjecture based on an empirical study on existing works.
Even though we cannot prove the conjecture at this time, 
we believe that \textsc{Gpt} conjecture will help researcher to have a deeper understanding 
of the nature of CFI problem and it will direct future research in this area.


\newpage
\bibliographystyle{IEEEtran}
\bibliography{main}
\end{document}